\ifx\onecol\undefined
\documentclass[journal,twocolumn]{IEEEtran}
\else 
\documentclass[12pt,draftclsnofoot,journal,onecolumn]{IEEEtran}
\fi
\usepackage{amsfonts}
\usepackage{amsmath,amssymb}
\usepackage{acronym}  
\usepackage{algorithmic}
\usepackage{algorithm}
\usepackage{bm}
\usepackage{bbm}
\usepackage{booktabs}
\usepackage{color, soul}
\usepackage{cite}
\usepackage{graphicx}
\usepackage{indentfirst}
\usepackage{lipsum}
\usepackage{setspace}
\usepackage{tikz}
\usetikzlibrary{arrows}
\usepackage{subfigure}
\usepackage[amsmath,thmmarks]{ntheorem}
\usepackage{theorem}
\usepackage{url}
\usepackage{array}
\usepackage{multirow}
\usepackage{caption}
\begin{document}

\title{FM-Receiver: A Foundation Model Enabled Unified Inner and Outer Neural Receiver Towards AI-Native Wireless Communications}

\author{Tianyue~Zheng, Chao~Jiang, and Linglong~Dai, {\textit{Fellow, IEEE}}
\thanks{
This work was funded in part by the National Science Fund for Distinguished Young Scholars (Grant No. 62325106), and in part by the National Key R\&D Program of China (No. 2023YFB811503). 
}
\thanks{All the authors are with the Department of Electronic Engineering, Tsinghua University, Beijing 100084, China, and also with the State Key Laboratory of Space Network and Communications, Tsinghua University, Beijing 100084, China (e-mails: \{zhengty22, jiangc24\}@mails.tsinghua.edu.cn, \{daill\}@tsinghua.edu.cn).
}}

\maketitle
\begin{abstract}
    With the development of artificial intelligence (AI) techniques, neural receivers, which apply AI to improve wireless receivers have been developed.
	However, most existing neural receivers apply deep learning only to the outer receiver while retaining conventional channel decoding for the inner receiver, which prevents joint optimization and makes it difficult to build efficient and unified AI-native receivers.
	To address this issue, we propose a foundation model (FM)-enabled unified neural receiver, FM-Receiver, that integrates the outer and inner receivers into a single AI-native framework, by leveraging the strong representation capability of FMs.
	Specifically, we introduce a grouped error correction code Transformer that performs symbol-level channel decoding, enabling seamless integration of the inner and outer receiver.
	Building on this, we illustrate the proposed FM-Receiver, that directly takes the received signals as input of FM and outputs the recovered transmitted bits.
	In addition, a three-stage configuration-adaptive pre-training strategy is designed to improve the generalization ability to diverse system configurations and scenarios.
	Extensive simulations show that the proposed FM-Receiver achieves better performance than baselines across different system configurations. It also demonstrates strong zero-shot generalization to unseen frequency bands and scenarios.
\end{abstract}
\vspace{-1mm}
\begin{IEEEkeywords}
    Neural receiver, foundation model, channel decoding, joint optimization
\end{IEEEkeywords}

\vspace{-3mm}
\section{Introduction} \label{sec-intro}
\vspace{-1mm}
Artificial intelligence (AI) is increasingly regarded as a transformative technology for wireless communications. It offers significant potential to enhance performance and intelligence, and gives rise to the emerging paradigm of AI radio access network (AI-RAN)~\cite{AI6G,6G2021}.
In the prevailing modular transceiver architecture, AI techniques have been applied to various individual physical-layer tasks, such as beam management, positioning, channel state information (CSI) feedback, and  channel decoding~\cite{CSIFEED,ECCT}.
By effectively capturing complex relationships in wireless channels, these data-driven methods have consistently outperformed conventional model-based approaches.
Recognizing these advantages, the 3rd Generation Partnership Project (3GPP) has actively promoted the integration of AI into 5G and 5G-Advanced standardization~\cite{3GPP_RP241862}.
Studies in Release 18 and Release 19 have shown performance gains from AI-based CSI compression, beam management, and positioning~\cite{stan}.

With the development of 6G, the evolution from AI-assisted air interface to a more AI-native air interface is expected~\cite{Standard}, which is defined as the seamless, intrinsic integration of AI technologies into the network architecture and operations. 
Among the key enablers of this vision, the neural receiver stands out as a promising approach.
Unlike the traditional modular signal processing blocks, the neural receivers employ a neural network capable of jointly performing multiple blocks of receiver (e.g. channel estimation, equalization, etc.)~\cite{MCMC,DM}.

\vspace{-3mm}
\subsection{Prior Works}
Various neural network architectures have been explored to implement neural receivers, with CNN-based and Transformer-based designs being the two prominent approaches.
Among CNN-based methods, a representative work is DeepRx~\cite{deeprx}, which proposed a fully convolutional neural network capable of jointly performing channel estimation, equalization, and demapping over an entire OFDM resource grid to directly generate log-likelihood ratios (LLRs) as the input of channel decoder.
Furthermore, NVIDIA extended the CNN-based framework from single-user to multi-user MIMO scenarios in~\cite{CGNN} based on Sionna platform, which integrates graph neural networks (GNNs) to effectively mitigate inter-user interference while supports a flexible number of users.
To validate this idea, NVIDIA further implements CNN-based methods in real world 5G NR systems~\cite{receiver_5g}, demonstrating their practicality and performance advantages under realistic 5G NR conditions.

More recently, Transformer architectures have been explored to enhance neural receivers by exploiting the distinguished feature of self-attention to capture long-range dependencies across time and frequency.
A Transformer-based neural receiver by Softband in~\cite{transformer}, which demonstrates strong performance in different use cases, highlighting the benefits of attention-based global modeling for practical wireless receivers.
To reduce the complexity of standard self-attention in Transformer,~\cite{axis} introduced an axial self-attention mechanism that factorizes attention operations along the temporal and spectral axes, substantially reducing computational cost while maintaining comparable performance.
Moreover,~\cite{DAT} proposed dual attention Transformer-based neural receiver to enhance modeling capacity by incorporating specialized attention mechanisms tailored to wireless channel characteristics.

Beyond the aforementioned architectural designs, numerous studies have further enhanced neural receiver performance from different perspectives.
To mitigate the difficulty of acquiring large-scale scenario-specific channel data,~\cite{CDM} proposed a conditional diffusion model-based framework that generates high-fidelity synthetic channel samples, thereby enabling effective data augmentation and improving neural receiver performance.
The authors in~\cite{Pilotless} developed a framework that jointly optimizes asymmetric modulation constellations and a neural receiver, allowing reliable pilotless communications through implicit channel estimation directly from data symbols.
Besides, to address performance degradation caused by channel distribution shifts during operation,~\cite{continue} proposed a zero-overhead continual learning framework that redesigns demodulation reference signals (DMRS) to simultaneously support signal detection and online model adaptation.
Moreover, ~\cite{QAT} introduced a quantization-aware training methodology that enables neural receivers efficiently deploy on resource-constrained hardware.

However, despite the advancements mentioned above, most neural receivers apply neural networks only to the outer receiver (i.e., channel estimation, equalization, and demapping), while retaining conventional channel decoding in the inner receiver. For example, the method proposed by NVIDIA in~\cite{CGNN} applied neural network to outer reciever, while employing belief propagation (BP) for channel decoding.
On the other hand, 
although channel decoding can also be realized by diverse deep-learning based methods~\cite{ECCT, BPdecoder},
these methods usually assume perfect LLRs with AWGN channel, neglecting the imperfect LLRs produced by the outer receiver. 
This separate design between the inner and outer receivers fails to achieve joint optimization of the receiver, 
resulting in performance degradation especially in challenging scenarios.
It also hinders the development of unified and efficient AI-native receiver architectures with reliable performance.

\vspace{-3mm}
\subsection{Our Contributions}
In this paper, we propose an FM-enabled unified neural receiver, FM-Receiver, to resolve the performance degradation, by leveraging the powerful representation learning capabilities of foundation models (FMs)\footnote{Simulation codes will be provided to reproduce the results in this paper after publication: \url{http://oa.ee.tsinghua.edu.cn/dailinglong/publications/publications.html}.}. 
It integrates the outer and inner receiver into a single AI-native framework. 
Through joint optimization of the entire receiver with diverse pretrained datasets, the proposed architecture delivers improved performance across different scenarios and system configurations.

Our contributions are summarized as follows.
\begin{itemize}
	\item Firstly, we propose an FM-enabled unified neural receiver  architecture, namely FM-Receiver, that directly takes the received signals as input of FM and outputs the recovered transmitted bits.
Unlike existing approaches that apply neural networks only to the outer receiver while retaining conventional BP decoding for the inner receiver, or focus only on AI-based channel decoding, our method unifies the outer and inner receivers using stacked Transformer blocks, and jointly optimizes the entire receiver.
	\item A key design of the proposed FM-Receiver is grouped error correction Transformer (G-ECCT), which achieves symbol level channel decoding instead of bit level channel decoding. Specifically, conventional AI-based channel decoders operate at the bit level, where each token corresponds to a single bit. It creates a mismatch with the outer receiver, whose minimum processing unit is a modulated symbol consisting of multiple bits. The mismatch not only hinders the seamless unification of the inner and outer receiver, but also leads to high computational complexity. In contrast, the proposed G-ECCT groups the $m$ bits of a modulated symbol into a single token. Thus, it can facilitate better integration of the outer and inner receiver while reducing computational complexity by a factor of $m^2$. To achieve the idea, group-based mask generation scheme, and specialized pre- and post-processing process are introduced.
	\item Moreover, we design a three-stage configuration-adaptive pre-training strategy to ensure the generalization capability of the proposed FM-Receiver in practical deployments.
Specifically, the outer receiver should adapt for different channel models, frequency bands, users number and velocities,
while the inner receiver requires simultaneous support of varying code lengths, code rates, and bits per symbol, which imposes higher demands on the generalization capability of  FM-Receiver.
Thus, we pre-train the model on large-scale datasets encompassing diverse scenarios, frequency bands, and modulation and coding schemes (MCSs). The three pretraining stages focus on the outer, inner, and joint inner-outer receiver optimization, respectively.
This enables FM-Receiver to effectively adapt to a wide range of system configurations.
	\item Finally, we conduct extensive simulations to validate the performance of the proposed FM-Receiver. The proposed method can achieve performance improvements across a wide range of system configurations, including varying numbers of users, MCSs, and numbers of subcarriers. 
	For example, for two-user CDL-A scenario, the proposed FM-Receiver achieves over 1.5 dB performance gain than existing methods for different MCSs and Eb/$\rm n_0$.
	Furthermore, it demonstrates strong zero-shot generalization capability under unseen scenarios and frequency bands. In addition, we analyze the scalability of the proposed method with respect to both pre-training dataset size and model scale, as well as its computational complexity and inference time.
\end{itemize}

\vspace{-3mm}
\subsection{Organization and Notation}
\subsubsection{Organization}
The rest of the paper is organized as follows.
Section~\ref{Sec_2} presents the system model, including the structure of the transmitter, the conventional receiver, and the neural receiver. Section~\ref{GECCT} introduces the proposed G-ECCT. In Section~\ref{proposed}, we integrate G-ECCT with the outer receiver to develop the FM-enabled uified neural receiver. Section~\ref{sec-re} provides extensive simulation results to demonstrate the performance and generalization capability of the proposed FM-Receiver. Finally, Section~\ref{sec-con} concludes the paper.

\subsubsection{Notation}
Bold lowercase letters denote vectors and bold uppercase letters denote matrices. For a vector $  \mathbf{a}  $ and a matrix $  \mathbf{A}  $, $  \mathbf{a}^H  $ and $  \mathbf{A}^H  $ denote the conjugate transpose.
$  \|\mathbf{a}\|_2  $ denotes the $  \ell_2  $-norm of a vector and $  \|\mathbf{A}\|_F  $ denotes the Frobenius norm of a matrix.
$  \mathbb{R}  $ and $  \mathbb{C}  $ denote the sets of real and complex numbers, respectively. $  \mathcal{CN}(\mu, \boldsymbol{\Sigma})  $ denotes the complex multivariate Gaussian distribution with mean $  \mu  $ and covariance matrix $  \boldsymbol{\Sigma}  $.
The operator $  \odot  $ denotes the element-wise (Hadamard) product. The functions $  \text{sign}(\cdot)  $ and $  \text{bin}(\cdot)  $ represent the element-wise sign function and the mapping from signs to binary values $  \{0,1\}  $, respectively.
$  \mathbf{I}  $ denotes the identity matrix and $  \sigma^2  $ denotes the noise variance.

\vspace{-4mm}
\section{System Model} \label{Sec_2}

\begin{figure*}
	\centering 
	\includegraphics[width= \linewidth]{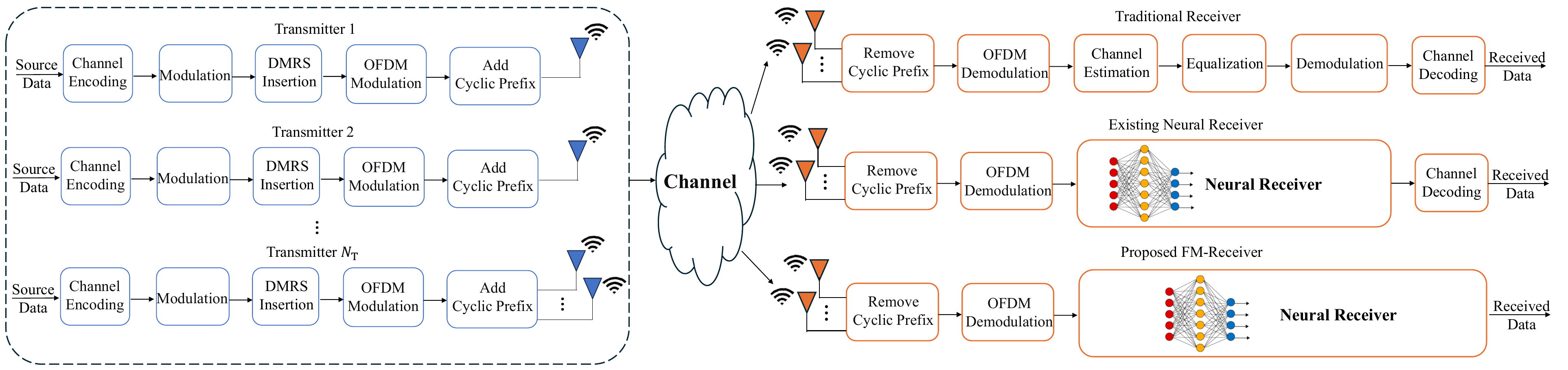}
	\caption{System model with traditional reciever, existing neural receiver, and proposed unified neural receiver.}
	\vspace{-3mm}
	\label{fig:system}
\end{figure*}

We consider a MU-MIMO system on the physical uplink shared channel (PUSCH), in which $N_{\rm T}$ users are simultaneously transmitted on the same physical resources to a base station (BS) equipped with $N_{\rm R}$ receive antennas~\footnote{Without loss of generality, we assume that each user is equivalent to one data stream.  For multi-antenna users transmitting multiple data streams, we treat each stream as a virtual user. This enables the proposed neural receiver to support multiple data streams per user by processing each stream independently. Thus, throughout the remainder of this paper, the number of users $N_{\rm T}$ is equivalent to the number of data streams.}. Uniform linear arrays (ULAs) with half-wavelength spacing between antenna elements are employed.
An orthogonal frequency division multiplexing (OFDM) scheme is adopted, where each slot consists of $N_F$ subcarriers and $N_S$ consecutive symbols, resulting in $V = N_FN_S$ time-frequency resource elements (REs) in a resource grid (RG) of size $N_F \times N_S$.
The considered communication system is illustrated in Fig.~\ref{fig:system}.
In this section, we sequentially present the structure of the transmitter and traditional receiver. Then basis of neural receiver is discussed.

\vspace{-4mm}
\subsection{Transmitter Architecture} \label{Sec_2_Subsec_1} 

As shown in Fig.~\ref{fig:system}, a random sequence of uniformly distributed information bits is generated at the transmitter. 
These bits are then encoded with a low-density parity check (LDPC) encoder, which adds redundancy to the input bitstream for robustness against channel errors. 
The resulting encoded bits are converted into complex-valued baseband symbols, and distributed over the available resource elements  allocated for data transmission. Specifically, we denote the encoded bits transmitted by the $n_{\rm T}$-th user on the $[n_S,n_F]$-th resource element as $\mathbf{b}_{n_S,n_F,n_{\rm T}} \in \{0,1\}^m$. $\mathbf{b}_{n_S,n_F,n_{\rm T}}$ is mapped to a complex symbol $\mathbf{x}_{n_S,n_F,n_{\rm T}} \in \mathbb{C}$ typically using a $2^m$ quadrature amplitude modulation (QAM) with Gray labeling.

Next, demodulation reference signal (DMRS) insertion process is performed. To be specific, except for the resource elements carrying data symbols, a subset of resource elements is reserved for DMRS.
DMRS symbols, serving as known pilots for channel estimation, are inserted at pre-defined positions in each slot.
In the considered system, DMRS is placed at OFDM symbols 2 and 11~\cite{5gnr}, with each transmission data stream assigned a dedicated DMRS port.
Therfore, the baseband modulated symbols for the  resource grid (RG) of size $N_F \times N_S$ can be written as
\begin{equation}
\mathbf{X}_{n_T} = 
\begin{bmatrix}
x_{1,1,n_{\rm T}} & \cdots & x_{1,N_S,n_{\rm T}} \\
\vdots & \ddots & \vdots \\
x_{N_F,1,n_{\rm T}} & \cdots & x_{N_F,N_S,n_{\rm T}}
\end{bmatrix}.
\end{equation}
Then MIMO precodig can be applied if a user is equipped with more than one transmit antennas.

Afterwards, the data of $n_{\rm T}$-th user is turned into an OFDM waveform by feeding $\mathbf{X}_{n_T}$ into an inverse fast Fourier transform (IFFT). Finally, the inter-symbol interference is mitigated by adding a cyclic prefix (CP) to the start of each OFDM symbol, and the waveform is transmitted by the channel.

\vspace{-4mm}
\subsection{Traditional Receiver Architecture} \label{Sec_2_Subsec_2}

After propagating through the channel, the BS antenna array simultaneously receives the transmitted symbols from $N_{\rm T}$ users. We assume OFDM with a sufficiently long CP duration is employed. Hence, after removing the CP and applying FFT to each individual OFDM symbol, the received signal $\mathbf{y}_{n_F,n_S} \in \mathbb{C}^{N_{\rm R}}$ of the $[n_S,n_F]$-th RE can be expressed as
\begin{equation}
	\mathbf{y}_{n_F,n_S} = \mathbf{H}_{n_F,n_S} \mathbf{x}_{n_F,n_S} + \mathbf{n}_{n_F,n_S},
\end{equation}
where $\mathbf{x}_{n_F,n_S} = [x_{n_F,n_S,1},\dots,x_{n_F,n_S,N_{\rm T}}]^\top \in \mathbb{C}^{N_{\rm T}}$ denotes the vector of transmitted baseband symbols, $\mathbf{H}_{n_F,n_S} \in \mathbb{C}^{N_{\rm R} \times N_{\rm T}}$ is the channel matrix, and $\mathbf{w}_{n_F,n_S} \sim \mathcal{CN}(0, \sigma^2 \mathbf{I}_{N_{\rm R}})$\ represents the complex-valued additive white Gaussian noise (AWGN) with noise power $\sigma^2$. For multi-antenna user, $\mathbf{H}_{n_F,n_S}$ denotes the effective channel matrix, involving both the precoding matrix at the transmitter and the physical channel matrix. 
The objective of the receiver is to  reconstruct the bits transmitted by the individual users from $\mathbf{y}_{n_F,n_S}$ which is corrupted by the noise and inter-user interference. The detailed process is summarized as follows.

Firstly, the reciever performs channel estimation using the received DMRS. Specifically, employing a least squares(LS) estimator, the channel on the RE $[n_F,n_S]$ carrying pilot symbols of user $n_{\rm T}$ is calculated as
\begin{equation}
	\hat{\mathbf{h}}_{n_F,n_S,n_{\rm T}} = \frac{x_{n_F,n_S,n_{\rm T}}^{p*}\mathbf{y}_{n_F,n_S}}{|x_{n_F,n_S,n_{\rm T}}^p|^2},
\end{equation}
where $x_{n_F,n_S,n_{\rm T}}^p \in \mathbb{C}$ denotes transmitted DMRS. 
Then linear interpolation is used to provide channel estimates for RE carrying the data symbols.
This will result in the entire channel estimate $\hat{\mathbf{H}}_{n_F,n_S}$. 
Other channel estimation methods, such as linear minimum mean square error (LMMSE), can also be applied to obtain  $\hat{\mathbf{H}}_{n_F,n_S}$. 

The next step is equalization which estimates the frequency domain signal using the estimated channel.
More precisely, an LMMSE equalizer is typically utilized in the receiver, which results in the equalizer output 
\begin{equation}
\hat{x}_{n_F,n_S} = \left( \hat{\mathbf{H}}_{n_F,n_S}^H \hat{\mathbf{H}}_{n_F,n_S} + \hat{\sigma}_n^2 \mathbf{I} \right)^{-1} \hat{\mathbf{H}}_{n_F,n_S}^H \mathbf{y}_{n_F,n_S},
\end{equation}
where $\hat{\sigma}_n^2$ is the estimated noise power and $\mathbf{I}$ is the identity matrix. 
The equalized symbols are next fed to the demapper, projecting the symbols to their corresponding constellation points. In this case, the bit-level LLRs of the $m$ bits $\mathbf{b}_{n_S,n_F,n_{\rm T}}$ corresponding to $\mathbf{x}_{n_S,n_F,n_{\rm T}}$ is obtained as
\begin{equation}
	LLR(\mathbf{b}_{n_S,n_F,n_{\rm T}}[i]) = \log \left( \frac{P(\mathbf{b}_{n_S,n_F,n_{\rm T}}[i]=0|\hat{\mathbf{x}}_{n_S,n_F,n_{\rm T}})}{P(\mathbf{b}_{n_S,n_F,n_{\rm T}}[i]=1|\hat{\mathbf{x}}_{n_S,n_F,n_{\rm T}})}\right),
\end{equation}
where $\mathbf{b}_{n_S,n_F,n_{\rm T}}[i]$ denotes the $i$-th bit in $\mathbf{b}_{n_S,n_F,n_{\rm T}}$, while $P(\cdot|\hat{\mathbf{x}}_{n_S,n_F,n_{\rm T}})$ is the conditional probability given the observed symbol $\hat{\mathbf{x}}_{n_S,n_F,n_{\rm T}}$.

Finally, the LLRs are passed to the channel decoder to recover the transmitted bit sequence $\mathbf{s}$.

\vspace{-3mm}
\subsection{Neural Receiver} \label{Sec_2_Subsec_3}
\cite{deeprx} first proposed the concept of neural receiver, which aims to directly operate on the received resource grid after FFT, and generate the LLRs for decoding process.
In other words, it effectively replaces the conventional modules of channel estimation, equalization, and demapping. 
Unlike traditional receivers, where each functional block is optimized independently, the neural receiver is trained in an end-to-end fashion. Specifically, the network is optimized by minimizing the binary cross-entropy (BCE)~\cite{bce} loss between the soft outputs (i.e. LLRs) produced by the neural receiver and the ground-truth encoded bits. Through this joint optimization, the neural receiver can achieve significant performance gains over conventional receivers.

However, current neural receivers typically stop at LLR generation, leaving channel decoding to traditional algorithms such as BP. To address this limitation, we propose a unified neural receiver, FM-Receiver, that incorporates decoding within the neural architecture.
This design involves the entire receiver  by taking the resource grid as input and directly recovering the transmitted bits. It is promising to further enhance the performance of neural receivers and contribute to AI-native physical layer implementations.
To realize the proposed FM-Receiver, we first develop a G-ECCT for decoding, which will be introduced in Section III. Building on this component, the overall architecture of FM-Receiver is presented in Section IV.

\vspace{-3mm}
\section{Proposed Grouped Error Correction Code Transformer (G-ECCT)} \label{GECCT}
To achieve a unified receiver architecture, we aim to implement the channel decoding component using a Transformer-based structure as well, enabling joint optimization with the outer receiver (channel estimation, equalization, and demapping).
Although~\cite{ECCT} has proposed an efficient Transformer-based decoder, this design has several limitations. Specifically, the decoder in~\cite{ECCT} performs bit-level channel decoding, where each token corresponds to a single bit. This creates a mismatch with the symbol-level received signals processed by the outer receiver, making seamless unification of the neural receivers challenging. Furthermore, the computational complexity of Transformer grows significantly as codeword length increases.

To address these issues, we propose a symbol-level decoder that supports multiple modulation schemes, where each token represents one modulated symbol (comprising $m$ bits). This design aligns directly with the received symbol-level signals. To provide a clear illustration to our proposed G-ECCT, this section first reviews the bit-level channel decoder presented in~\cite{ECCT} and then introduces the proposed symbol-level decoder.

\vspace{-3mm}
\subsection{Review of Error Correction Code Transformer (ECCT) }\label{Sec_3_Subsec_1}
Forward error correction (FEC) is a fundamental technique in communication systems to reduce the bit error rate (BER) over noisy channels. We consider a binary linear block code defined by a generator matrix $\mathbf{G} \in \mathbb{R}^{k \times n}$ and a parity-check matrix $\mathbf{H} \in \mathbb{R}^{(n-k) \times n}$ satisfying $\mathbf{G}\mathbf{H}^T = \mathbf{0}$ over the binary field. A message vector $\mathbf{s} \in \{0,1\}^k$ is encoded into a codeword $\mathbf{b} \in \{0,1\}^n$ via $\mathbf{b} = \mathbf{s} \cdot \mathbf{G}$, which satisfies $\mathbf{H}\mathbf{b} = \mathbf{0}$.
At the receiver, the decoder takes the LLRs of the received signals as input and aims to recover a soft approximation $\hat{\mathbf{b}}$ of the transmitted codeword $\mathbf{b}$. An illustration of this channel decoding framework is presented in Fig.~\ref{fig:system}.

\subsubsection{Pre- and post- processing}
ECCT adopts pre-processing and post-processing of~\cite{prepost} to avoid overfitting and achieve MMSE decoding.
In the pre-processing stage, the received LLRs, denoted as $\mathbf{l} \in \mathbb{R}^n$, are first transformed into a vector of a dimensionality $2n-k$ suitable for feature extraction. Specifically, the absolute values of the LLRs and the syndrome information are concatenated as follows:
\begin{equation}
\tilde{\mathbf{l}} = \bigl[ |\mathbf{l}|, \, s(\mathbf{l}) \bigr],
\end{equation}
where $[\cdot, \cdot]$ denotes vector concatenation, $|\mathbf{l}|$ represents the element-wise absolute value of the LLR vector, and $s(\mathbf{l}) = \mathbf{H} \cdot \text{bin}(\text{sign}(\mathbf{l})) \in \{0,1\}^{n-k}$ is the binary syndrome computed from the hard decisions of the LLRs.
Obtaining $\tilde{\mathbf{l}}$, positional reliability encoding is performed, which projects $\tilde{\mathbf{l}}$ to a high $d$ dimensional embedding $\mathbf{X}_{\rm in}$. Then $\mathbf{X}_{\rm in}$ is utilized as the input of the Transformer.

\cite{ECCT} models the channel decoding problem as $\mathbf{l}=(1-2\mathbf{b}) \cdot \mathbf{z}$, where $\mathbf{z}$ is the random multiplicative noise. The Transformer-based decoder is trained to predict the noise vector, and recover the transmitted codeword from the noise vector. 
Thus, in the post-processing stage, the output of the Transformer $\mathbf{X}_{\rm out} \in \mathbb{R}^{(2n-k) \times d}$ is projected to $\hat{\mathbf{z}} \in \mathbb{R}^{n}$. 
This prediction is then combined with the input LLR vector $\mathbf{l}$ to recover the estimated codeword. Specifically, the final hard-decision output is obtained as
\begin{equation}\label{eq-bits}
\hat{\mathbf{x}} = \text{bin}\bigl(\text{sign}({\mathbf{l}} \odot \hat{\mathbf{z}})\bigr),
\end{equation}
where $\odot$ denotes element-wise multiplication, $\text{sign}(\cdot)$ is the element-wise sign function, and $\text{bin}(\cdot)$ maps the signs back to binary values $\{0,1\}$.

\subsubsection{Code-aware self attention based Transformer}
Obtaining $\mathbf{X}_{\rm in}$, ECCT leverages the self-attention mechanism of the Transformer to analyze the received bits and detect the errors. 
Furthermore, in order to improve the performance of decoding, ECCT integrates code structure as fundamental domain knowledge into attention mechanisms. To be specific, since not every bit is necessarily related to all the others, employing full attention, where each token attends to all other tokens, is sub-optimal. Therefore, a mask $g(\mathbf{H}) : \{0,1\}^{(n-k) \times k} \rightarrow \{-\infty, 0\}^{(2n-k) \times (2n-k)}$ is introduced according to $\mathbf{H}$ and applied to the self-attention mechanism to indicate the relationships among the bits. The construction algorithm of the mask is summarized in Algorithm~\ref{mask_ecct}.
The mask is initialized as the identity matrix. For each row $i$ of the parity-check matrix $\mathbf{H}$, we unmask all pairwise positions corresponding to the ones in that row, as these bits are directly connected through the parity-check equation. Additionally, we unmask the connections between these bit positions and the corresponding syndrome bit at position $n+i$, since they jointly define the parity-check constraints. Thus, the code-aware self attention mechanism can be written as 
\begin{equation}\label{mask_att}
\begin{aligned}
&\text{Attn}(\mathbf{W}^O, \mathbf{W}^K, \mathbf{W}^V) 
=  \\ &\text{Softmax}\left( \frac{ (\mathbf{X}\mathbf{W}^O)(\mathbf{X}\mathbf{W}^K)^T + g(\mathbf{H})}{ \sqrt{d_k} } \right)
\cdot (\mathbf{X}\mathbf{W}^V),
\end{aligned}
\end{equation}
where $\mathbf{W}^O, \mathbf{W}^K, \mathbf{W}^V$ are learnable matrices.

\begin{figure*}
	\centering 
	\subfigure[]{
		\includegraphics[width=0.31 \linewidth]{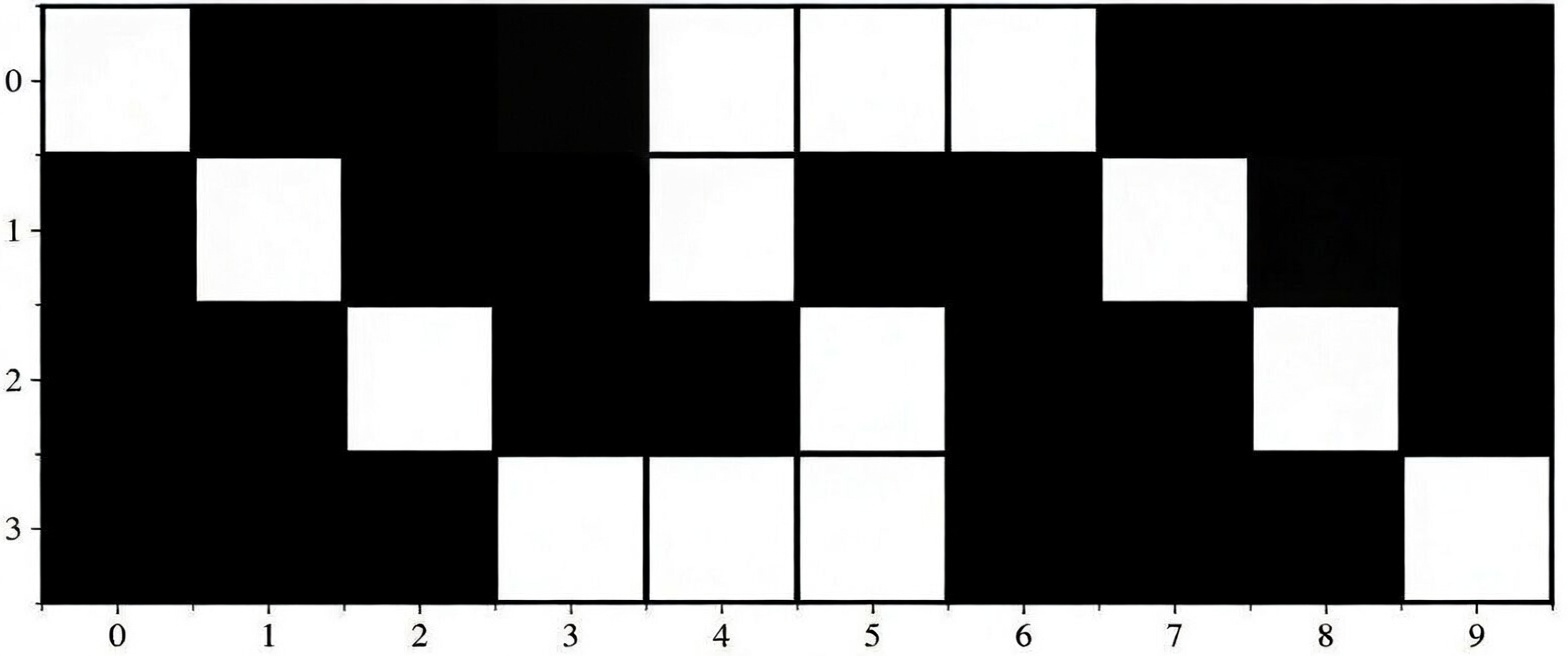}} 
        \subfigure[]{
		\includegraphics[width=0.3 \linewidth]{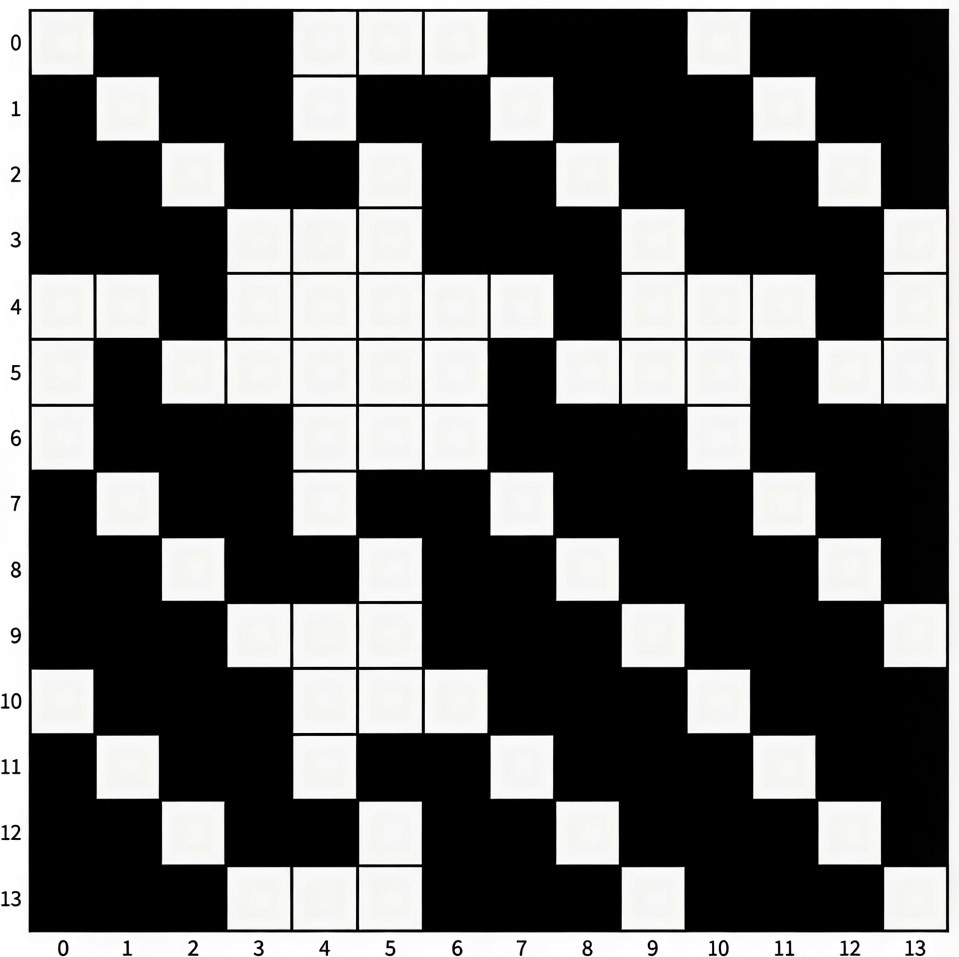}}
	\subfigure[]{
		\includegraphics[width=0.25 \linewidth]{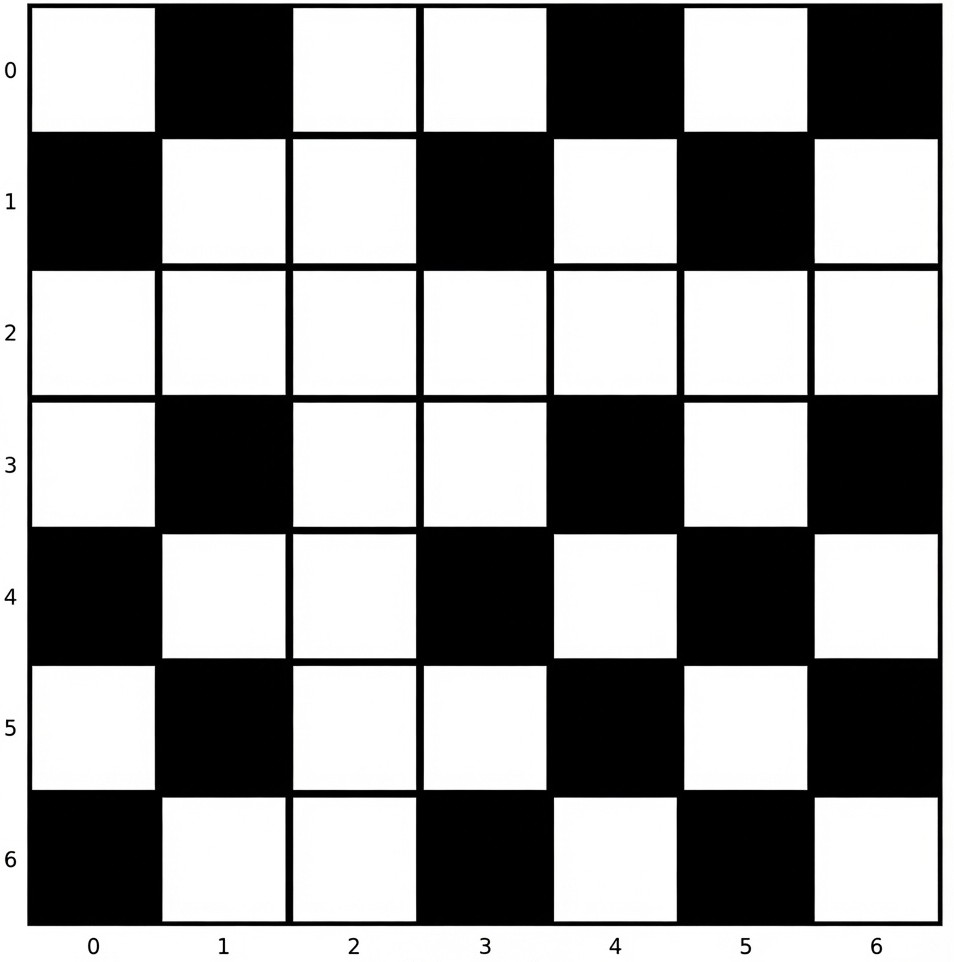}}
	\caption{Illustration with LDPC ($n=10,k=6$): (a) parity check matrix $\mathbf{H}$; (b) mask generated in ECCT; (c) mask generated in proposed G-ECCT. White squares represent 1 and black squares represent 0.}
	\label{pic_mask}
	\vspace{-3mm}
\end{figure*}

For enhanced clarity, a (10, 6) LDPC code is taken as an example to provide a visual depiction of parity check matrix $\mathbf{H} \in \{0,1\}^{4 \times 10}$ and the generated mask $g(\mathbf{H})$ in Fig.~\ref{pic_mask}. Here, the matrix $\mathbf{H}$ is defined as
\begin{equation}
\mathbf{H} = 
\begin{bmatrix}
1 & 0 & 0 & 0 & 1 & 1 & 1 & 0 & 0 & 0 \\
0 & 1 & 0 & 0 & 1 & 0 & 0 & 1 & 0 & 0 \\
0 & 0 & 1 & 0 & 0 & 1 & 0 & 0 & 1 & 0 \\
0 & 0 & 0 & 1 & 1 & 1 & 0 & 0 & 0 & 1
\end{bmatrix}.
\end{equation}
Finally, the output of the Transformer can be written as $\mathbf{X}_{\rm out}$.

\begin{algorithm}[t]
\caption{Mask Construction Pseudo Code of ECCT}\label{mask_ecct}
\begin{algorithmic}[1] 
\STATE \textbf{function} $g(\mathbf{H})$
\STATE \quad $l, n \gets \mathbf{H}.\text{shape}$
\STATE \quad $k \gets n-l$
\STATE \quad $\text{mask} \gets \text{eye}(2n-k)$
\STATE \quad \textbf{for} $i = 0$ \textbf{to} $n-k-1$ \textbf{do}
\STATE \quad \quad $\text{idx} \gets \text{where}(\mathbf{H}[i]==1)$
\STATE \quad \quad \textbf{for} $j \in \text{idx}$ \textbf{do}
\STATE \quad \quad \quad $\text{mask}[n+i,j] \gets \text{mask}[j,n+i] \gets 1$
\STATE \quad \quad \quad \textbf{for} $k \in \text{idx}$ \textbf{do}
\STATE \quad \quad \quad \quad $\text{mask}[j,k] \gets \text{mask}[k,j] \gets 1$
\STATE \quad \quad \textbf{end for}
\STATE \quad \quad \textbf{end for}
\STATE \quad \textbf{end for}
\STATE \quad \textbf{return} $-\infty(-\text{mask})$
\STATE \textbf{end function}
\end{algorithmic}
\end{algorithm}

\vspace{-3mm}
\subsection{Grouped Error Correction Code Transformer} \label{Sec_3_Subsec_2}
As shown previously, ECCT operates as a bit-level decoder by embedding the LLR of each bit as a token. To better accommodate symbol-level received signals and enable a unified receiver architecture, this section introduces the proposed G-ECCT, transforming bit-level decoding into symbol-level decoding.
Specifically, as detailed in Section~\ref{Sec_2}, the receiver obtains modulated symbols, each consisting of $m$ bits. Correspondingly, each token output by the outer receiver carries the LLRs of $m$ bits.
Accordingly, the core idea of G-ECCT is to treat the LLRs of every $m$ bits as a group, project them into one token, and perform decoding at the symbol level, enabling seamless integration of the outer and inner receiver.

Implementing G-ECCT presents two key challenges. First, since users may adopt different modulation schemes, the number of bits per symbol ($m$) varies. Thus, G-ECCT needs to flexibly support different values of $m$. Second, the mask used in ECCT to involve domain knowledge cannot be directly applied. Consequently, a dedicated masking mechanism must be designed for G-ECCT to effectively capture the relationships among tokens. In the following, we elaborate on how the proposed G-ECCT addresses the aforementioned challenges.

\subsubsection{Pre- and post- processing}
To support symbol-level decoding and flexible adaptation to different modulation orders, we design dedicated pre- and post-processing modules for G-ECCT.
Let $\mathbf{l}_g \in \mathbb{R}^{n_1 \times m}$ denote the LLR vector output by the outer receiver, where each row corresponds to the $m$ LLRs of one modulated symbol and $n_1$ is the number of symbols. 
Then, we also group every $m$ bits of $\mathbf{s}$ to form the grouped syndrome $\mathbf{s}_g \in \{0,1\}^{n_2 \times m}$, where $n_2 = (n-k) / m$.
Next, we concatenate the LLRs and the corresponding syndrome:
\begin{equation}\label{preprocessing}
\tilde{\mathbf{l}}_g = [|\mathbf{l}_g|, \mathbf{s}_g] \in \mathbb{R}^{n_0 \times m},
\end{equation}
where $n_0 = n_1+n_2$.

To enable a unified input format for different modulation schemes (i.e., varying $m$), we introduce a maximum modulation order $m_{\max}$. We pad $\tilde{\mathbf{l}}_g$ with zeros in the feature dimension to obtain the final input tensor of shape $n_0 \times m_{\max}$. This padding ensures that the network can process inputs with a fixed feature dimension regardless of the actual $m$ used in the modulation scheme.

After passing through the Transformer-based decoder, the network outputs the predicted noise vector $\mathbf{z} \in \mathbb{R}^{n_1 \times m_{\max}}$. In the post-processing stage, we first extract the relevant portion according to the actual modulation order $m$ to obtain the grouped noise $\mathbf{z}_g \in \mathbb{R}^{n_1 \times m}$. In this case, we recover the estimated codeword $\hat{\mathbf{x}}$ using the same procedure as in ECCT.

This pre- and post-processing design allows G-ECCT to efficiently handle symbol-level inputs while maintaining compatibility with various modulation schemes and seamless integration with the outer receiver.

\subsubsection{Group-based mask generation}
To incorporate the code structure into G-ECCT, we design a dedicated attention mask tailored for symbol-level grouped decoding. The overall procedure is presented in Algorithm~\ref{alg:mask_g_ecct}.
Specifically, we first construct the bit-level self-attention mask for the original ECCT decoder using the parity check matrix $\mathbf{H}$ (as detailed in Algorithm~\ref{mask_ecct}). 
Then, after grouping the bits into tokens of size $m$, we generate the grouped mask $\text{mask}_g$. 
For any two groups $i$ and $j$, if there exists at least one pair of bits where one belongs to group $i$ and the other to group $j$ that are connected in the bit-level mask, we set $\text{mask}_g[i,j] = 1$ (and symmetrically $\text{mask}_g[j,i] = 1$).
This grouped mask $\text{mask}_g$ effectively preserves the essential structural relationships among symbols while enabling efficient symbol-level processing in the Transformer architecture.

\begin{algorithm}[t]
\caption{Mask Construction Pseudo Code of G-ECCT}
\label{alg:mask_g_ecct}
\begin{algorithmic}[1] 
\STATE \textbf{function} $g_g(\mathbf{H}, m)$
\STATE \quad Compute $\text{mask}$ according to Algorithm~\ref{mask_ecct}
\STATE \quad $l, n \gets \mathbf{H}.\text{shape}$, $k \gets n-l$
\STATE \quad $n_0 \gets \lceil 2n-k / m \rceil$ 
\STATE \quad $\text{mask}_g \gets \mathbf{0}^{n_0 \times n_0}$
\STATE \quad \textbf{for} $i = 0$ \textbf{to} $n_0-1$ \textbf{do}
\STATE \quad \quad \textbf{for} $j = i$ \textbf{to} $n_0-1$ \textbf{do}
\STATE \quad \quad \quad connected $\gets$ \textbf{False}
\STATE \quad \quad \quad \textbf{for} $p = i \cdot m$ \textbf{to} $(i+1)\cdot m-1$ \textbf{do}
\STATE \quad \quad \quad \quad \textbf{for} $q = j \cdot m$ \textbf{to} $(j+1)\cdot m-1$ \textbf{do}
\STATE \quad \quad \quad \quad \quad \textbf{if} $\text{mask}[p,q] = 1$ \textbf{then}
\STATE \quad \quad \quad \quad \quad \quad connected $\gets$ \textbf{True}
\STATE \quad \quad \quad \quad \quad \quad \textbf{break}
\STATE \quad \quad \quad \quad \quad \textbf{end if}
\STATE \quad \quad \quad \quad \textbf{end for}
\STATE \quad \quad \quad \quad \textbf{if} connected \textbf{then} \textbf{break} \textbf{end if}
\STATE \quad \quad \quad \textbf{end for}
\STATE \quad \quad \quad \textbf{if} connected \textbf{then}
\STATE \quad \quad \quad \quad $\text{mask}_g[i,j] \gets 1$, $\text{mask}_g[j,i] \gets 1$
\STATE \quad \quad \quad \textbf{end if}
\STATE \quad \quad \textbf{end for}
\STATE \quad \textbf{end for}
\STATE \quad \textbf{return} $-\infty(-\text{mask}_g)$
\STATE \textbf{end function}
\end{algorithmic}
\end{algorithm}

It is worth noting that when the number of subcarriers of RG varies, the input length $n_0$ of G-ECCT also changes accordingly. Thanks to the Transformer's inherent ability to handle variable-length sequences, no extra design is required. We can simply feed inputs with different code lengths and their corresponding masks into the model for joint training.
After developing G-ECCT, we integrate G-ECCT with the outer receiver and present the FM-enabled unified neural receiver in the following section.

\vspace{-3mm}
\section{Proposed FM-Enabled Unified Neural Receiver} \label{proposed}
This section presents the proposed FM-enabled unified inner and outer neural receiver, FM-Receiver.
The method aims to implement the entire receiver using neural networks while supporting flexible adaptation to varying system configurations, including different numbers of users, frequency bands, numbers of sub-carriers, and MCSs.
We first introduce the overall framework of FM-Receiver. Then, in Section~\ref{Sec_4_Subsec_2}, we detail the network architecture and explain how it achieves unified neural receivers. Finally, we describe the pre-training strategy designed to enable the receiver to effectively handle diverse user and numbers, frequency bands, and MCSs, etc..

\vspace{-3mm}
\subsection{Overall Framework}\label{Sec_4_Subsec_1}
\begin{figure*}
	\centering 
	\includegraphics[width= \linewidth]{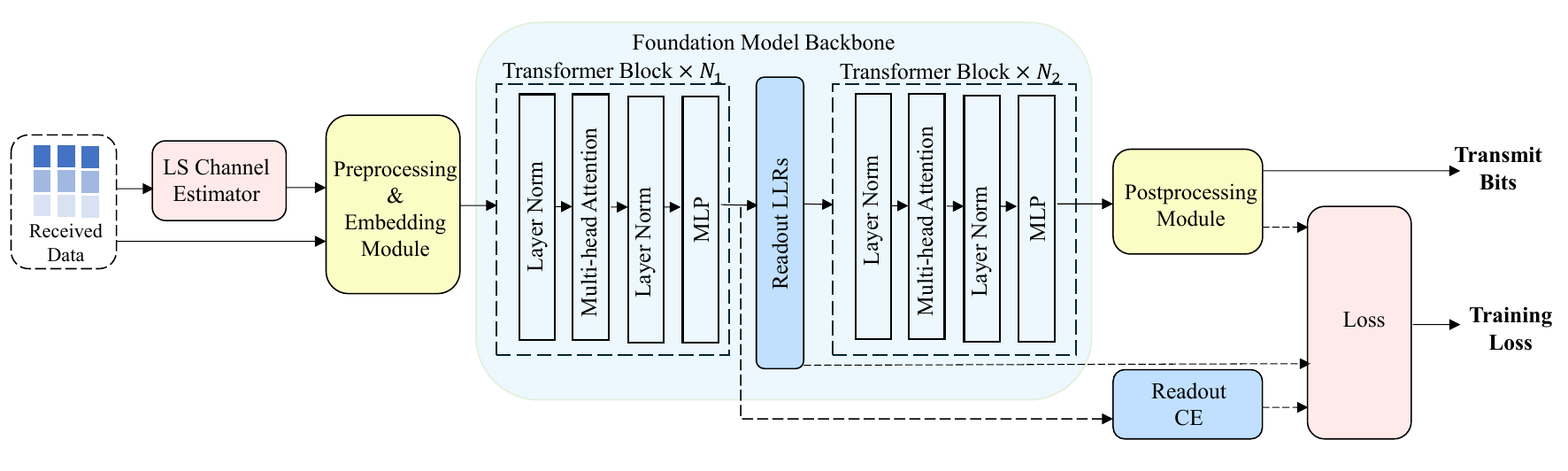}
	\caption{Overall framework of proposed unified neural receiver.}
	\label{fig:framework}
	\vspace{-3mm}
\end{figure*}

The framework of the proposed FM-enabled unified neural receiver is illustrated in Fig.~\ref{fig:framework}.
The receiver takes the received signal $\mathbf{Y}$ as input and directly outputs the estimated transmitted bits, encompassing channel estimation, equalization, demodulation, and channel decoding within a unified neural architecture.

As shown in the figure, the received signal $\mathbf{Y} \in \mathbb{C}^{N_F \times N_S \times N_{\rm R}}$ first undergoes LS channel estimation to obtain the channel coefficient estimate $  \hat{\mathbf{H}} \in \mathbb{C}^{N_{\rm T}\times N_F \times N_S \times N_{\rm R}}$.
Both $\mathbf{Y}$ and $\hat{\mathbf{H}}$ are then fed into the network. It should be noted that while it is not strictly necessary to provide the raw channel estimate as input, it facilitates the learning process by offering a bootstrap for subsequent channel estimation and equalization tasks, which would otherwise require multiple neural layers to approximate~\cite{deeprx}.

After appropriate pre-processing and embedding that align with the input format of the foundation model backbone, the concatenated features are processed by foundation model. In the backbone, both the outer receiver and G-ECCT are implemented by stacked Transformer blocks. 
Specifically, full-attention Transformer blocks perform frequency-domain signal processing to suppress noise and multi-user interference, producing LLRs. Subsequently, G-ECCT employs masked-attention Transformer blocks to learn the multiplicative noise from these LLRs. Finally, post-processing modules recover the transmitted bits from the refined soft information. 
This unified architecture enables joint optimization across all receiver modules while maintaining high flexibility and strong performance.

\vspace{-3mm}
\subsection{Network Structure}\label{Sec_4_Subsec_2}
This subsection describes the network structure of the proposed FM-enabled unified neural receiver.
The network takes  the complex-valued received post-FFT signal $\mathbf{Y}$, a three-dimensional tensor of shape $N_F \times N_S \times N_{\rm R}$, as input.
Then, the LS channel estimator, detailed in Section~\ref{Sec_2_Subsec_2}, produces the channel estimate $\mathbf{H}_{n_{\rm T}}$ for each user, which also has the shape $N_F \times N_S \times N_{\rm R}$.

\textbf{Preprocessing \& Embedding Module}. 
These two complex valued tensors are jointly processed by the preprocessing \& embedding module.
The module first converts $\mathbf{Y}$ and $\mathbf{H}_{n_{\rm T}}$ into real-valued tensors by concatenating the real and imaginary parts along the feature dimension. The resulting tensors are then concatenated along the same dimension, producing an initial input tensor $\mathbf{X}_{init}$ of shape $ N_F \times N_S \times 4N_{\rm R}$ for the $n_{\rm T}$-th user.
Then $\mathbf{X}_{init}$ is normalized and passed through the embdedding module. This module applies a stack of convolutional layers interleaved with ReLU activations to extract hierarchical time-frequency features, followed by a final projection layer that maps the features to the model dimension $d_s$.
The output is rearranged into a tensor of shape $N_FN_S \times d_s$.
Then position embeddings are added to provide sequential information, obtaining the prepared input $\mathbf{X}_{pre}$ to the FM backbone.

\textbf{FM backbone}. 
The FM backbone consists of a total of $N = N_1 + N_2$ Transformer blocks stacked sequentially. 
The first $N_1$ blocks employ full self-attention and are dedicated to frequency-domain signal processing in the outer receiver. These blocks progressively extract and integrate features to suppress noise and multi-user interference while generating LLRs.
The subsequent $N_2$ blocks utilize masked self-attention to realize symbol-level decoding through G-ECCT, as described in Section~\ref{Sec_3_Subsec_2}.

Each of the first $N_1$ Transformer blocks follows the standard architecture with two sub-layers: a multi-head self-attention module and a position-wise feed-forward multi-Layer perceptron (MLP) network.
Let $\mathbf{X}_{\rm FM}^{I(l)}$ denote the input to the $l$-th block ($l = 1, \dots, N_1$). By definition, the input to the first block is the embedded feature tensor $\mathbf{X}_{pre}$, while subsequent blocks receive the output of the preceding block, i.e., $\mathbf{X}_{\rm FM}^{I(l)} = \mathbf{X}_{\rm FM}^{O(l-1)}$.
Layer normalization (LN) is first applied independently across the feature dimension for each token and each sample in the batch:
\begin{equation}
	\mathbf{X}_{\rm FM}^{{\rm ln1}(l)}= \text{LayerNorm}(\mathbf{X}_{\rm FM}^{I(l)}).
\end{equation}
The normalized features are then processed by the multi-head self-attention module with a residual connection: 
\begin{equation}
	\mathbf{X}_{\rm FM}^{{\rm att}(l)} = \text{ATT}(\mathbf{X}_{\rm FM}^{{\rm ln1}(l)}) + \mathbf{X}_{\rm FM}^{{\rm ln1}(l)},
\end{equation}
where $\text{ATT}(\cdot)$ denotes multi-head self-attention, which can be written as
\begin{equation}
\begin{aligned}
&\text{Attn}(\mathbf{W}^O, \mathbf{W}^K, \mathbf{W}^V) 
=  \\ &\text{Softmax}\left( \frac{ (\mathbf{X}\mathbf{W}^O)(\mathbf{X}\mathbf{W}^K)^T}{ \sqrt{d_k} } \right)
\cdot (\mathbf{X}\mathbf{W}^V).
\end{aligned}
\end{equation}
A second layer normalization is subsequently applied: 
\begin{equation}
	\mathbf{X}_{\rm FM}^{{\rm ln2}(l)}= \text{LayerNorm}(\mathbf{X}_{\rm FM}^{{\rm att}(l)}).
\end{equation}
The normalized features are passed through the MLP module with another residual connection:
\begin{equation}
	\mathbf{X}_{\rm FM}^{O(l)} = \text{MLP}(\mathbf{X}_{\rm FM}^{{\rm ln2}(l)}) + \mathbf{X}_{\rm FM}^{{\rm ln2}(l)},
\end{equation}
where $\text{MLP}(\cdot)$ denotes the position-wise feed-forward network. The output $\mathbf{X}_{\rm FM}^{O(l)}$ of the $l$-th block serves as the input to the $(l+1)$-th block. 

After successively processing through all $N_1$ full-attention Transformer blocks, a lightweight read-out module is applied to $\mathbf{X}_{\rm FM}^{O(N_1)}$ to produce the LLRs of the resource elements. 
Specifically, a linear projection layer maps the feature dimension $d_s$ to $m_{\rm max}$, the maximum number of bits supported per modulated symbol. This yields an LLR tensor of shape $N_FN_S \times m_{\rm max}$ containing soft information for up to $m_{\rm max}$ bits at every resource element:
\begin{equation}
	\hat{\mathbf{l}}_g = \text{MLP}(\mathbf{X}_{\rm FM}^{O(N_1)}).
\end{equation}
To accommodate different modulation schemes with varying $m$, the network always projects to the maximum dimension $m_{\rm max}$. For any given modulation order, only the first $m$ LLR values are retained while the remaining entries are discarded. This unified projection strategy enables the outer receiver to flexibly support a wide range of modulation schemes without requiring structural modifications.

Obtaining the grouped LLRs $\hat{\mathbf{l}}_g$ from the outer receiver, symbol-level decoding is performed by G-ECCT, which consists of the remaining $N_2$ masked-attention Transformer blocks.
Following the procedure described in Section~\ref{GECCT}, the syndrome is first computed. 
The LLRs and syndrome are then preprocessed (refer to Eq.~\eqref{preprocessing}) and projected into the backbone feature dimension $d_s$, yielding the input feature map $\mathbf{X}_{\rm FM}^{I(N_1+1)}$ to the $(N_1+1)$-th Transformer block.
The subsequent $N_2$ blocks share the same architecture as the preceding full-attention blocks, with the only difference being that multi-head self-attention is replaced by masked self-attention (Eq.~\eqref{mask_att}) to exploit the code structure. 
The blocks are stacked sequentially, and the output of the final  block is denoted $\mathbf{X}_{\rm FM}^{O(N)}$.

\textbf{Postprocessing Module}. 
Finally, the postprocessing module recovers the estimated transmitted codeword from $\mathbf{X}_{\rm FM}^{O(N)}$.
Specifically, a linear projection layer is first applied to $\mathbf{X}_{\rm FM}^{O(N)}$ to predict the noise vector $\hat{\mathbf{z}}_g$. The predicted noise is then used, together with the received LLRs, to reconstruct the estimated codeword $\hat{\mathbf{x}}$ following the recovery procedure defined in Eq.~\eqref{eq-bits}. 
The resulting hard decisions of $\hat{\mathbf{x}}$ constitute the final output of the FM-enabled unified neural receiver.

Furthermore, we also introduce a read-out CE module. This auxiliary read-out layer takes the final state representation $\mathbf{X}_{\rm FM}^{O(N_1)}$ as input and produces a refined channel estimate $\hat{\mathbf{H}}_{n_{\rm T}}$ for each user. The module is implemented as a lightweight MLP that processes independently at each resource element:
\begin{equation}
	\hat{\mathbf{H}}_{n_{\rm T}} = \text{MLP}(\mathbf{X}_{\rm FM}^{O(N_1)}).
\end{equation}
The read-out CE module generates more accurate channel estimates by leveraging the rich features already extracted by the FM backbone. 
Besides, when supervised by an additional mean-squared-error (MSE) loss term between the predicted and ground-truth channels, it is observed to significantly improve training convergence and overall performance stability~\cite{receiver_5g}.
This auxiliary supervision will be illustrated later in Section.~\ref{Sec_4_Subsec_3}.

\vspace{-4mm}
\subsection{Pre-Training}\label{Sec_4_Subsec_3}
Leveraging the strong generalization capability of FMs, we pretrain the proposed FM-enabled unified neural receiver across multiple heterogeneous datasets that cover diverse scenarios and system configurations.
The objective is to enable the model to adapt to varying numbers of subcarriers, frequency bands, channel models, user numbers and velocities, as well as different modulation and coding schemes, while allowing direct application to unseen scenarios and configurations.
The pre-training procedure consists of three stages, which are described in detail below.

\textbf{Stage 1: Pretrain of outer receiver.}
First, we pretrain the outer receiver portion of the network (i.e., the first $N_1$ full-attention Transformer blocks together with preprocessing \& embedding module, the readout LLRs module, and readout CE module). 
The objective is to learn robust frequency-domain processing that effectively mitigates multi-user interference and channel fading effects.
For each training mini-batch, the number of active users $n_{\rm T}$ (where $1 \leq n_{\rm T} \leq N_{\rm T}$) is randomly sampled. Each  UE transmits randomly generated payload bits that are independently encoded and modulated.

In stage 1, the network takes the received post-FFT signal $\mathbf{Y}$ as input. After processing through the $N_1$ full-attention blocks, the readout LLR module produces LLR estimates $\hat{\mathbf{l}}_g$, while the readout CE module outputs refined channel estimates $\hat{\mathbf{H}}_{n_{\rm T}}$.
The training loss consists of two terms. The first is the BCE loss between the predicted LLRs and the ground-truth bit labels:
\begin{equation}
	\mathcal{L}_{\rm BCE} = -\frac{1}{B} \sum_{b=1}^{B} \sum_{i=1}^{N_b} \Bigl[ b_{b,i} \log \sigma(\hat{\mathbf{l}}_{b,i}) + (1 - b_{b,i}) \log \bigl(1 - \sigma(\hat{\mathbf{l}}_{b,i})\bigr) \Bigr],
\end{equation}
where $B$ is the batch size, $N_b$ is the total number of bits in the $b$-th sample, $b_{b,i} \in \{0,1\}$ denotes the ground-truth bit, and $\sigma(\cdot)$ is the sigmoid function. The second term is the MSE loss on the channel estimates, which supervises the readout CE module: 
\begin{equation}
	\mathcal{L}_{\rm MSE} = \frac{1}{B} \sum_{b=1}^{B} \bigl\| \hat{\mathbf{H}}_b - \mathbf{H}_b \bigr\|_F^2,
\end{equation}
where $\mathbf{H}_b$ is the ground-truth channel realization for the $b$-th sample. 
This auxiliary supervision helps the network not only yield accurate channel estimates but also improve training convergence without increasing inference complexity.
The total training loss is then given by
\begin{equation}
	\mathcal{L}_1 = \mathcal{L}_{\rm BCE} + \gamma \mathcal{L}_{\rm MSE},
\end{equation}
where the hyperparameter $\gamma > 0$ controls the relative contribution of the channel estimation loss.

\textbf{Stage 2: Pretrain of G-ECCT.}
In Stage 2, we pre-train the G-ECCT portion (i.e., $N_2$ masked-attention Transformer blocks, and postprocessing module) while keeping the parameters of the outer receiver fixed. 
Importantly, instead of using the LLRs produced by the Stage 1 outer receiver (which still contain errors and bias), we generate clean LLRs directly from an AWGN channel.
This design enables G-ECCT to converge rapidly and avoids overfitting to the imperfect LLR distribution of the outer receiver.
The decoders trained on AWGN channels have demonstrated basic generalization ability on fading channels without retraining~\cite{AWGN,UECCT}.

The training objective follows the multiplicative noise prediction framework in~\cite{ECCT}.
Let $\mathbf{l}_g$ denote the grouped LLR vector of a codeword. We aim to predict the binary multiplicative noise $\tilde{\mathbf{z}} = \operatorname{bin}(\mathbf{l}_g \odot (1-2\mathbf{b}))$.
The loss for a single grouped codeword is defined as the binary cross-entropy:
\begin{equation}
	\mathcal{L}_{\rm de} = -\frac{1}{B} \sum_{b=1}^{B} \sum_{i=1}^{N_b} \tilde{z}_{b,i} \log \sigma\bigl( \hat{z}_{b,i}\bigr) + (1 - \tilde{z}_{b,i}) \log \bigl(1 - \sigma(\hat{z}_{b,i}) \bigr),
\end{equation}
where $\hat{z}_{b,i}$ is the $i$-th element of the $b$-th sample. The estimated codeword is recovered as~\eqref{eq-bits}.
After Stage 2, G-ECCT acquires the capability to correct multiplicative noise under clean LLR conditions, providing a solid initialization for the subsequent joint training stage.

\textbf{Stage 3: Joint training of the unified neural reciever.}
In Stage 3, we jointly fine-tune the entire FM-enabled unified neural receiver. All parameters are updated end-to-end using diverse datasets. This joint optimization allows the outer receiver and G-ECCT to adapt to each other, further improving overall performance.
The training loss is the weighted sum of the three loss terms introduced in the previous stages: 
\begin{equation}
	\mathcal{L}_3 = \mathcal{L}_{\rm MSE} + \lambda_1 \mathcal{L}_{\rm BCE} + \lambda_2 \mathcal{L}_{\rm de},
\end{equation}
where $\lambda_1$ and $\lambda_2$ control the relative importance of the different terms.

\vspace{-3mm}
\section{Simulation Results} \label{sec-re}
This section presents comprehensive simulation results to evaluate the performance of the proposed FM-enabled unified neural receiver.
We first describe the datasets used for training and evaluation, followed by the detailed simulation setups, including network architecture, pre-training configurations, baseline methods, and performance metrics.
We then assess the proposed FM-Receiver under different MCSs and varying numbers of users, ect..
The zero-shot generalization capability to unseen scenarios and system configurations, as well as the scaling behavior with increasing model size, are further investigated.
Finally, we analyze the computational complexity and inference efficiency of the proposed method.

\vspace{-3mm}
\subsection{Datasets}\label{Sec_5_Subsec_1}

In this subsection, we illustrate the datasets we have constructed to train and evaluate the neural receiver.
For our experiments, we focus on the PUSCH scenario described in Section~\ref{Sec_2}, and utilize the Sionna~\cite{sionna} link-level simulator to model the PUSCH link compliant with the 3GPP standards.
A total of 15 datasets are simulated, among which the first 12 datasets are used for pre-training (indexed from D1 to D12) and the last three datasets are used for evaluation (indexed from T1 to T3).
The detailed simulation configurations of each dataset are shown in Table~\ref{tab:scenarios}.

To fully pretrain the FM and leverage its capacity to handle diverse scenarios and configurations, we employ 12 different datasets. These datasets cover a wide range of scenarios and user speed, and involve three different frequency bands and three subcarrier numbers. Moreover, each dataset supports varying numbers of users and MCS configurations. Specifically, the number of users is randomly selected from 1 to 4. For each user, the modulation scheme is randomly chosen from QPSK, 16-QAM, and 64-QAM, while the LDPC code rate is randomly selected from the set \{679/1024, 553/1024, 517/1024\}.
The three datasets T1-T3 are designed to comprehensively evaluate the proposed model. T1 employs scenarios similar to the training data to assess the model's multi-dataset learning capability as well as its robustness to variations in MCS, user number, and subcarrier number. In contrast, T2 and T3 focus on evaluating the model's zero-shot generalization to entirely new frequency bands and channel models.

Besides, for each dataset, the base station is equipped with 16 antennas, while each user is equipped with a single-antenna receiver. The antenna spacing is half-wavelength at the center frequency, and the subcarrier spacing is 15 kHz.

\begin{table*}[htbp]
\centering
\caption{Configurations of the Constructed Datasets.}
\label{tab:scenarios}
\begin{tabular}{@{}cccccc@{}}
\toprule
\textbf{Dataset} & \textbf{Channel Model} & \textbf{$f_c$ (GHz)} & \textbf{$N_F$} & \textbf{Delay Spread (ns)} & \textbf{User Speed (km/h)} \\
\midrule
D1  & CDL-A & 3.5  & 12 & 35  & [0, 30]  \\
D2  & CDL-A & 4.9  & 4  & 20  & [0, 30]  \\
D3  & CDL-A & 7.0  & 4  & 25  & [0, 30]  \\
D4  & CDL-C & 3.5  & 12 & 180 & [5, 70]  \\
D5  & CDL-C & 4.9  & 8  & 220 & [5, 70]  \\
D6  & CDL-C & 7.0  & 4  & 280 & [5, 70]  \\
D7  & CDL-D & 3.5  & 4  & 8   & [0, 120] \\
D8  & CDL-D & 4.9  & 4  & 12  & [0, 120] \\
D9  & CDL-D & 7.0  & 12 & 20  & [0, 120] \\
D10 & CDL-E & 3.5  & 4  & 75  & [3, 90]  \\
D11 & CDL-E & 4.9  & 8  & 50  & [3, 90]  \\
D12 & CDL-E & 7.0  & 12 & 80  & [3, 90]  \\
\midrule
T1 & CDL-A & 3.5  & 8  & 20  & [0, 30] \\
T2 & CDL-A & 28.0 & 4  & 35  & [0, 30] \\
T3 & CDL-B & 4.9  & 4  & 60  & [0, 60] \\
\bottomrule
\end{tabular}
\end{table*}

\vspace{-3mm}
\subsection{Simulation Setups}\label{Sec_5_Subsec_2}

\subsubsection{Network and Pre-training Settings}

\begin{table}[htbp]
\centering
\caption{Network parameters of the proposed model with different sizes.}
\label{tab:modelparas}
{\small
\begin{tabular}{cccc}
\toprule
\textbf{Model} & 
\begin{tabular}[c]{@{}c@{}}
\textbf{Feature} \\ 
\textbf{Dimension $d_s$}
\end{tabular} & 
\begin{tabular}[c]{@{}c@{}}
\textbf{Head} \\ 
\textbf{Number}
\end{tabular} & 
\textbf{Parameters} \\
\midrule
little & 128 & 8 & 2.84M \\
small & 256  & 8 & 10.16M \\
base  & 512  & 8 & 38.98M \\
large & 768  & 8 & 86.66M \\
\bottomrule
\end{tabular}}
\end{table}

To examine the effect of model scale on performance, we evaluate models with different sizes, ranging from 2.84M to 86.66M parameters.
The detailed configurations of these models are summarized in Table~\ref{tab:modelparas}.
For models of different sizes, we fix the number of Transformer blocks as $N_1 = N_2 = 6$.

\begin{table}[htbp]
\centering
\caption{Training configurations for the three pre-training stages.}
\label{tab:training_config}
{\small
\begin{tabular}{c|c|c|c}
\toprule
\textbf{Parameter} & \textbf{Stage 1} & \textbf{Stage 2} & \textbf{Stage 3} \\
\midrule
Optimizer& \multicolumn{3}{c}{AdamW (weight decay = 0.0001)} \\
Learning rate & $1.0 \times 10^{-4}$ & $1.0 \times 10^{-3}$ & $5.0 \times 10^{-5}$ \\
Batch size& 256  & 256  & 128  \\
Training steps& 200k & 5000k & 500k \\
\bottomrule
\end{tabular}}
\vspace{-3mm}
\end{table}

All training and inference of the proposed model are conducted on four NVIDIA GeForce RTX 4090 24GB GPUs. 
The training hyperparameters differ across the three stages, with the detailed settings summarized in Table~\ref{tab:training_config}.
During training, scenarios, MCSs, and the number of users are randomly sampled (as described in Section~\ref{Sec_4_Subsec_3}) to form each mini-batch.
In Stages 1 and 3, the Eb/$\rm n_0$ is uniformly sampled from the range [-4.0, 8.0] dB. In Stage 2, a range of [2.0, 8.0] dB is adopted, and a cosine learning rate decay scheduler is employed with an initial learning rate of 0.001.

\subsubsection{Baselines}
To comprehensively evaluate the proposed FM-enabled unified neural receiver, we compare it with both conventional and learning-based receivers as baselines.

\textbf{Conventional baselines}.
We consider two traditional receiver pipelines that employ BP decoding. 
The first baseline uses LS channel estimation followed by LMMSE equalization (denoted as ``LS+LMMSE+BP" in the simulation results). The second baseline replaces LS estimation with LMMSE channel estimation while retaining LMMSE equalization (denoted as ``LMMSE+LMMSE+BP" in the simulation results).
Both pipelines serve as strong classical references with relatively low computational complexity.

\textbf{Learning-based Baselines}.
For learning-based methods, existing methods mainly consider two neural architectures: CNN-based~\cite{CGNN,deeprx} and Transformer-based~\cite{transformer,transformer_axis}. 
For the CNN-based baseline, we adopt the convolutional neural network architecture proposed in~\cite{CGNN} (denoted as ``CNN+BP" in the simulation results).
This architecture leverages convolutional layers to exploit the time-frequency correlation of the wireless channel and employs a graph neural network (GNN) to effectively handle multi-user interference. We also implement conventional BP decoding.
For the Transformer-based baseline, we employ the same outer receiver architecture (i.e., the first $N_1$ full-attention blocks) as in our proposed method, to ensure a fair comparison with our proposed method. But we substitute the G-ECCT component with standard BP decoding (denoted as ``Transformer+BP" in the simulation results). 
This design allows us to conveniently evaluate the performance gain brought by unifing the outer receiver and channel decoding, while maintaining architectural consistency. 

\subsubsection{Performance Metrics}
We primarily adopt bit error rate (BER) as the main performance metric throughout this section to evaluate the reliability of different receiver architectures. 
All results are averaged over multiple independent channel realizations to ensure statistical reliability.

\vspace{-3mm}
\subsection{Performance Evaluation}\label{Sec_5_Subsec_3}

\subsubsection{Performance under System Configurations}

\begin{figure*}
	\centering 
	\subfigure[]{
		\includegraphics[width=0.30 \linewidth]{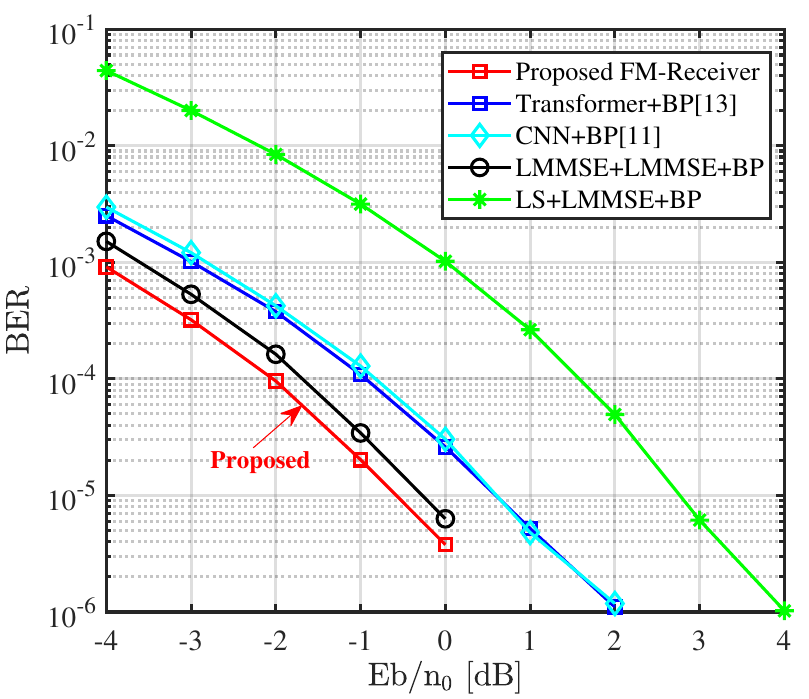}} 
        \subfigure[]{
		\includegraphics[width=0.32 \linewidth]{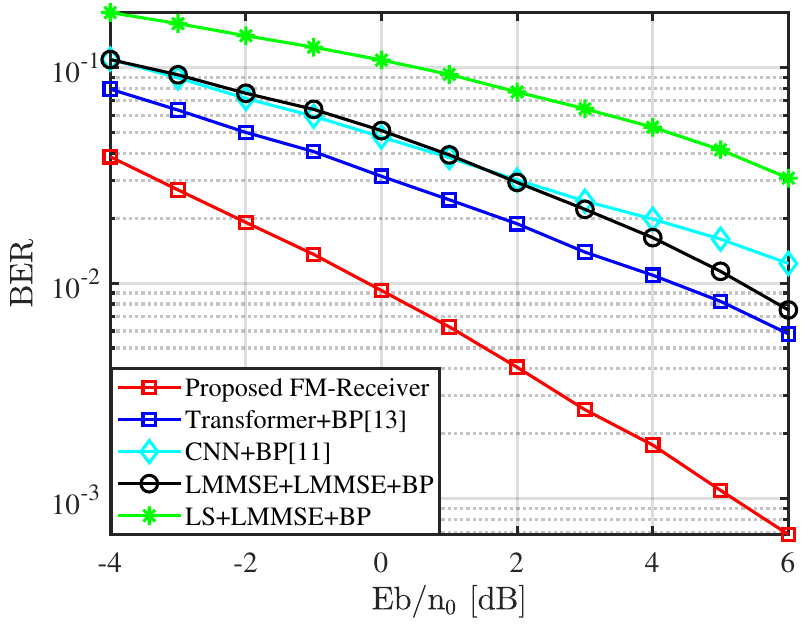}}
	\subfigure[]{
		\includegraphics[width=0.32 \linewidth]{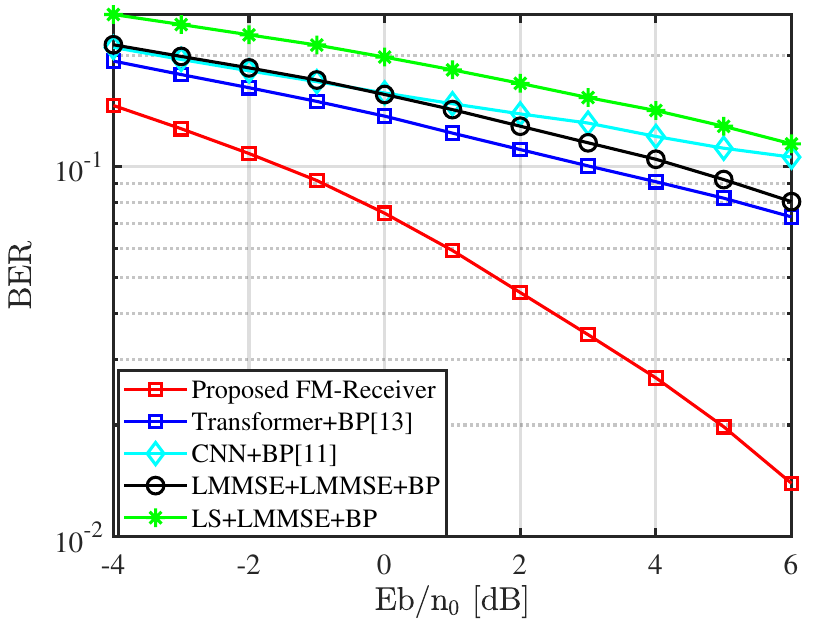}}
	\caption{Performance of different users: (a) single user; (b) 2 users; (c) 4 users.}
	\label{pic_user}
	\vspace{-3mm}
\end{figure*}

We first evaluate the performance of the proposed FM-enabled unified neural receiver under various system configurations using the T1 dataset.
Fig.~\ref{pic_user} presents the BER performance of different receivers under varying numbers of users. 
In these experiments, we fix the model size to the base configuration and adopt QPSK modulation with a code rate of 679/1024.
The number of active data streams is set to 1, 2, and 4 in Fig.~\ref{pic_user}(a), (b), and (c), respectively.
As shown in the figure, the proposed method consistently achieves the best performance across all user configurations.

In relatively simple single-user case (Fig.~\ref{pic_user}(a)), 
where existing methods already deliver strong performance,
the proposed method achieves only marginal gains over the baselines:
0.3 dB at BER = $10^{-5}$ performance gain compared to the ``LMMSE+LMMSE+BP" method and 0.8-1 dB over the Transformer-based baseline.
As the number of users grows, the performance advantage of the proposed receiver over the baselines becomes increasingly significant.
In the more challenging 4-user scenario (Fig. 5(c)), the proposed method demonstrates a substantial advantage, maintaining a gap of over 2 dB across the entire Eb/$\rm n_0$ range.
This trend indicates that the proposed FM-enabled unified neural receiver can more effectively suppress inter-user interference and combat noise in challenging multi-user scenarios. 
By jointly optimizing the outer receiver and the G-ECCT through end-to-end training, the proposed architecture is able to exploit the strong feature learning capability of FMs efficiently in challenging scenarios.

\begin{figure*}
	\centering 
	\subfigure[]{
		\includegraphics[width=0.31 \linewidth]{figures/plot_mcs0.pdf}} 
        \subfigure[]{
		\includegraphics[width=0.31 \linewidth]{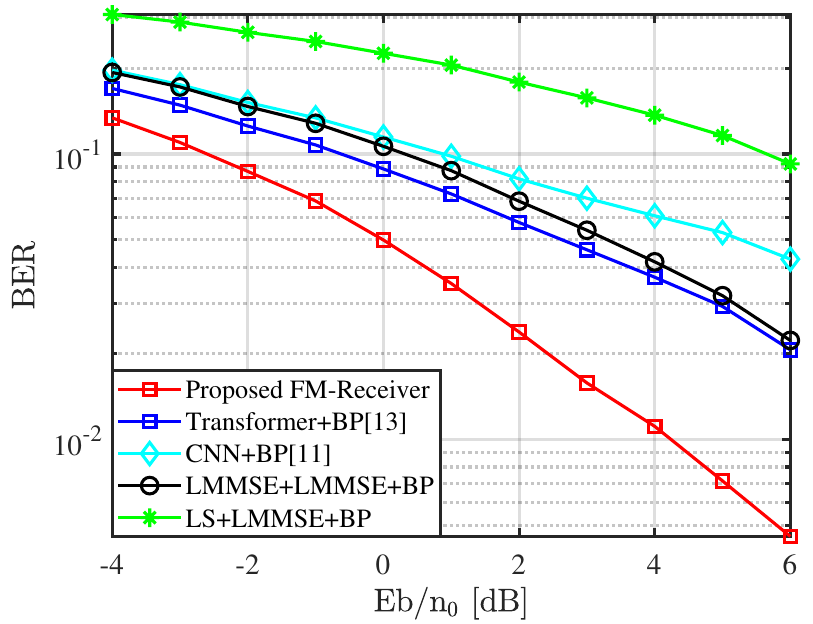}}
	\subfigure[]{
		\includegraphics[width=0.31 \linewidth]{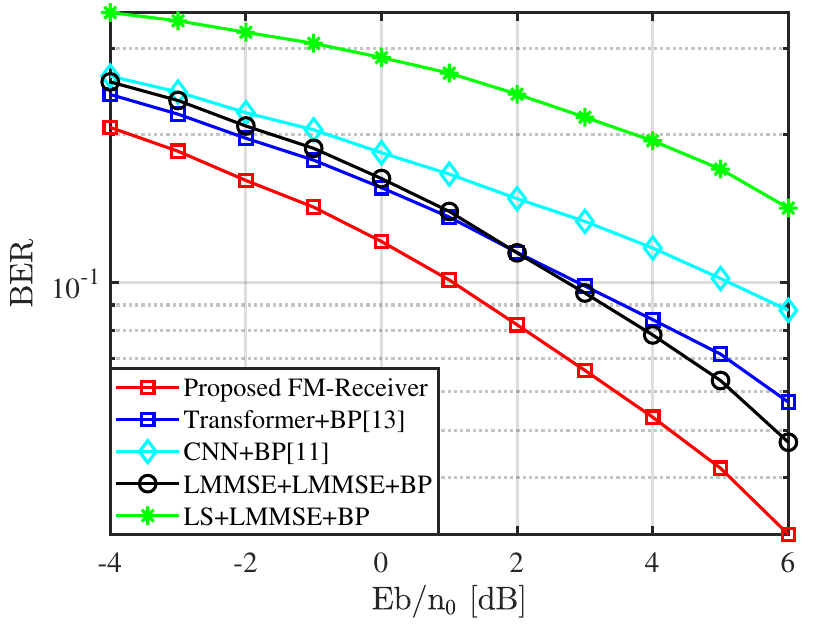}}
	\caption{Performance of different MCSs: (a) QPSK modulation, code rate = 679 / 1024; (b) 16-QAM modulation, code rate = 553 / 1024; (c) 64-QAM modulation, code rate = 517 / 1024.}
	\label{pic_mcs}
	\vspace{-3mm}
\end{figure*}

Fig.~\ref{pic_mcs} shows the BER performance of receivers under different MCSs.
In this set of experiments, we fix the model size to the base configuration and set the number of data streams to 2.
As shown in Fig.~\ref{pic_mcs}, the proposed method consistently achieves the best performance across all tested MCSs. 
In particular, it outperforms all baseline methods, including both conventional receivers and learning-based methods. 
This demonstrates the effectiveness of the proposed unified architecture that jointly optimizes the outer receiver and G-ECCT.
Compared with Transformer-based baseline, the performance gains of the proposed method primarily stem from two aspects.
First, the G-ECCT module, which employs masked-attention Transformer blocks for symbol-level decoding, provides performance benefits over conventional BP decoding.
Second, the end-to-end joint optimization of the outer receiver and G-ECCT enables better coordination between channel estimation/equalization and channel decoding, further enhancing overall reliability.

As shown in Fig.~\ref{pic_mcs}(a), under QPSK modulation, the proposed method achieves a substantial gain compared with the best traditional baseline (LMMSE-LMMSE-BP) and Transformer-based baseline (over 3 dB at BER $= 10^{-2}$).
In contrast, as the modulation order increases to 16-QAM and 64-QAM (Fig.~\ref{pic_mcs}(b) and (c)), the performance gap between the proposed FM-Receiver and existing methods becomes relatively smaller.
This is primarily because, as the code length increases, the performance of BP decoding approaches optimal, thereby diminishing the relative advantage of G-ECCT over it.
Nevertheless, even under the more challenging 64-QAM cases, the proposed receiver maintains a clear margin over baselines (over 1.5 dB), due to the advantage of joint optimization.
These results confirm that the proposed unified neural receiver not only delivers superior performance but also exhibits strong robustness across a wide range of modulation schemes.

\subsubsection{Zero-shot Generalization Ability}

\begin{figure}
	\centering 
	\includegraphics[width= 0.85 \linewidth]{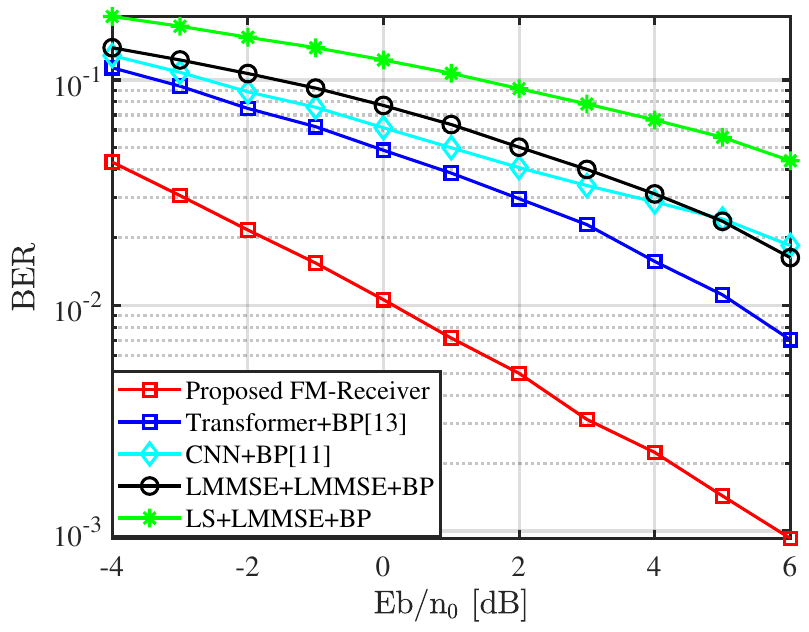}
	\caption{Generalization capability to a new frequency band.}
	\label{fig:gene_28}
	\vspace{-3mm}
\end{figure}

To evaluate the generalization capability of the proposed FM-enabled unified neural receiver, we first conduct cross-frequency band experiments using the T2 dataset.
The model is trained on data from 3.5 GHz, 4.9 GHz, and 7.0 GHz, and directly tested on a new frequency band at 28 GHz without any fine-tuning.
In these experiments, we use the base model size with 2 users, QPSK modulation, and code rate 679/1024.
As shown in Fig.~\ref{fig:gene_28}, the proposed method maintains strong performance in the unseen 28 GHz frequency band and consistently outperforms all baselines across the entire Eb/$\rm n_0$ range.

Furthermore, the performance gap between the proposed method and other learning-based baselines becomes even larger compared to the in-band results in Fig.~\ref{pic_mcs}(a).
For instance, the performance gain of the proposed method over the ``Transformer+BP" baseline is approximately 1 dB larger in the 28 GHz band than that observed in the original frequency band (Fig.~\ref{pic_mcs}(a)).
A similar trend can also be observed when compared with the CNN-based baseline.
These results suggest that the proposed FM-Receiver exhibits strong generalization capability when deployed in unseen frequency bands.

\begin{figure}
	\centering 
	\includegraphics[width= 0.85 \linewidth]{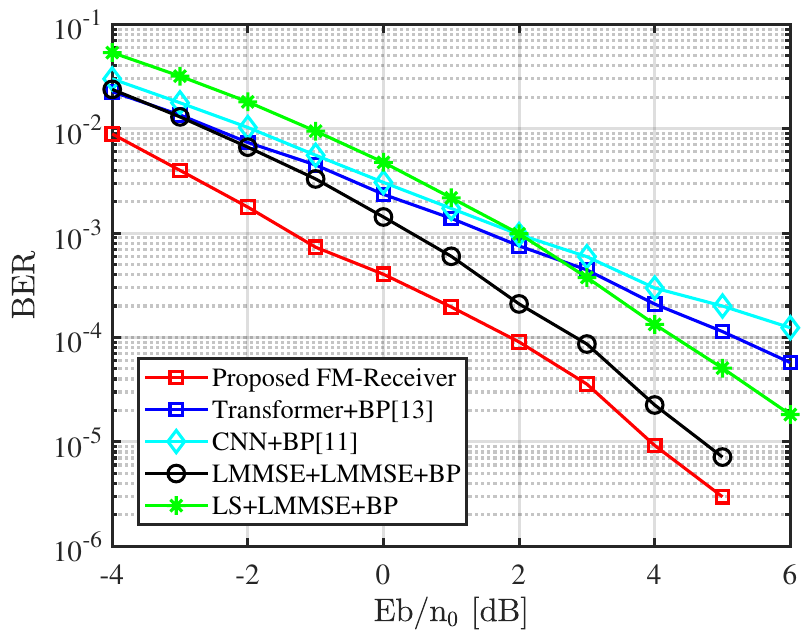}
	\caption{Generalization capability to a new channel model.}
	\label{fig:gene_cldb}
	\vspace{-3mm}
\end{figure}

Fig.~\ref{fig:gene_cldb} further evaluates the zero-shot generalization capability of the proposed FM-Receiver when tested on an unseen channel model.
As shown in the figure, both the ``Transformer+BP" and ``CNN+BP" baselines exhibit noticeable performance degradation.
This degradation can be attributed to two main factors. First, performing only the task of the outer receiver is relatively simple, neural networks tend to overfit the training distribution.
Second, these conventional neural receivers have fewer parameters than the proposed architecture, resulting in weaker generalization ability.
Consequently, their performance falls below that of the conventional LMMSE channel estimation with LMMSE equalization baseline, and at high Eb/$\rm n_0$, even approaches the performance of the simpler ``LS+LMMSE" receiver.
In contrast, the proposed FM-enabled unified neural receiver maintains strong performance under the unseen channel model and consistently outperforms the ``LMMSE+LMMSE+BP" baseline by approximately 1 dB across the entire Eb/$\rm n_0$ range.
This result demonstrates that, benefiting from large model capacity and extensive pre-training across diverse scenarios, the proposed method possesses superior generalization ability. It can effectively adapt to new channel models without any retraining, highlighting the advantage of the proposed unified FM-Receiver over conventional neural receiver designs.

\subsubsection{Scaling Ability}

\begin{table}[htbp]
\centering
\caption{The performance of proposed method evaluated across different model sizes. }
\label{tab:model_size}
{\small
\begin{tabular}{@{}ccccc@{}}
\toprule
\textbf{Dataset} & \textbf{Little} & \textbf{Small} & \textbf{Base} & \textbf{Large}\\
\midrule
T1  & $1.94 e^{-2}$ & $1.60 e^{-2}$ & \underline{$9.24 e^{-3}$} &  $\bf 8.77 e^{-3}$ \\
T2  & $2.11 e^{-2}$ & $1.40 e^{-2}$ &$\bf 1.04 e^{-2}$ &  \underline{$1.06 e^{-2}$}  \\
T3  & $1.29 e^{-3}$ & $6.44 e^{-4}$ & \underline{$4.60 e^{-4}$} & $\bf 2.38 e^{-4}$ \\
\bottomrule
\end{tabular}}
\end{table}

Scaling analysis is essential for foundation models, as it reveals how model capacity and pre-training data scale jointly influence both in-distribution learning performance and zero-shot generalization.
In this work, we investigate the scaling behavior of the proposed FM-enabled unified neural receiver by evaluating four model sizes (Little, Small, Base, and Large) and models with different pre-training datasets with different characteristics.

As shown in Table~\ref{tab:model_size}, increasing the model size generally leads to performance improvements across most datasets.
The best results are highlighted in \textbf{bold}, while the second-best results are \underline{underlined}.
We set user number as two, Eb/$\rm n_0$ as 0 dB, and MCS as QPSK with code rate = 679 / 1024.
On the T1 and T3 datasets, the ``Large" model size achieves the best results. 
This improvement can be attributed to the enhanced representation capacity of larger models, which enables them to capture more complex spatial-frequency patterns and interference/noise structures in the received signals. 
On the T2 datasets, the ``Base" and ``Large" model size achieve comparable performance.
This observation suggests that, once the model reaches a certain capacity, its generalization performance becomes constrained by the scale and diversity of the pre-training data, potentially leading to overfitting on specific dataset distributions.

\begin{table}[htbp]
\centering
\caption{The performance of proposed method evaluated across different pretrained datasets.}
\label{tab:dataset}
{\small
\begin{tabular}{@{}cccc@{}}
\toprule
\textbf{Dataset} & \textbf{D2,D5,D8,D11} & \textbf{D4-D9} & \textbf{D1-D12} \\
\midrule
T1  & $\underline{1.19 e^{-2}}$ & $2.38 e^{-2}$ & \textbf{$\bf 9.24 e^{-3}$} \\
T2  & $\underline{2.71 e^{-2}}$ & $3.25 e^{-2}$ & \textbf{$\bf 1.04 e^{-2}$}\\
T3  & $1.05 e^{-3}$ & $\underline{4.94 e^{-4}}$ & \textbf{$\bf 4.60 e^{-4}$} \\
\bottomrule
\end{tabular}}
\end{table}

To further investigate the impact of pre-training data scale and diversity on model generalization, we evaluate the proposed receiver when trained on different combinations of pre-training datasets while keeping the model size fixed at the base configuration. 
All models are tested under the same setting with 2 users at Eb/$\rm n_0$ = 0 dB using QPSK modulation with a code rate of 679/1024. The results are summarized in Table~\ref{tab:dataset}.
The best results are highlighted in \textbf{bold}, while the second-best results are \underline{underlined}.
When trained on the largest and most diverse dataset combination (D1-D12), the model achieves the best performance on three tasks.
Notably, on the more challenging T2 dataset (28 GHz), increasing the pre-training data diversity brings particularly large gains, reducing the BER from $2.71 e^{-2}$ (D2,D5,D8,D11) and $3.25 e^{-2}$ (D4-D9) to $1.04 e^{-2}$ (D1-D12).
This demonstrates that broader coverage of channel models and frequency bands during pre-training enhances the model's ability to generalize to unseen system configurations.

Although the D4-D9 dataset combination contains a larger number of datasets, it underperforms the more diverse D2,D5,D8,D11 combination on both the T1 and T2 test sets.
This suggests that simply increasing data volume is insufficient; the diversity and distribution of the pre-training data also play a critical role in achieving robust zero-shot generalization. Overall, these results indicate that both the scale and diversity of pre-training datasets are essential for improving the generalization capability of the proposed FM-based receiver.

\vspace{-3mm}
\subsection{Computational Costs}\label{Sec_5_Subsec_4}
\begin{table*}[htbp]
\centering
\caption{Computational costs including the number of parameters, and inference time per batch of different models (batch size = 8).}
\label{tab:cost}
{\small
\begin{tabular}{@{}cccccc@{}}
\toprule
\textbf{Model} & \textbf{LS+LMMSE+BP} & \textbf{LMMSE+LMMSE+BP}&\textbf{CNN+BP} & \textbf{Transformer+BP} & \textbf{Proposed}\\
\midrule
Parameters (M) & / & / & 1.11 & 19.90 & 38.98 \\
Inference time (ms) & 11.304 & 49.980 & 10.549 & 12.758 & 5.511 \\
\bottomrule
\end{tabular}}
\vspace{-3mm}
\end{table*}

To evaluate the computational efficiency of the proposed receiver, we measure the parameter number and inference time per batch on a single NVIDIA RTX 4090 GPU. The experiments are conducted with 2 users at Eb/$\rm n_0$ = 0 dB using QPSK modulation (code rate 679/1024) on the T3 dataset, with a batch size of 8. The detailed computational costs, including the number of parameters and inference latency, are summarized in Table~\ref{tab:cost}.
As shown in the table, the proposed method has the largest number of parameters (38.98M) among all compared methods. However, it achieves the lowest inference time of only 5.511 ms per batch, which is faster than all other baselines. In particular, it is more than 2x faster than the ``Transformer+BP" baseline (12.758 ms) and nearly 9x faster than the ``LMMSE+LMMSE+BP" receiver (49.980 ms).

This result can be attributed to the high inference latency of conventional BP decoding, which requires a large number of iterations, especially under low Eb/$\rm n_0$ conditions. In contrast, the proposed G-ECCT module performs symbol-level decoding in a single forward pass through the masked-attention Transformer blocks, thereby avoiding the iterative overhead of traditional BP decoding. 
Although the proposed model contains more parameters, its feed-forward architecture makes it suitable for practical deployment where both performance and real-time processing are required.

\section{Conclusions} \label{sec-con}
In this paper, we propose an FM-enabled unified neural receiver architecture, namely FM-Receiver, that integrates the outer and inner receivers, achieving the entire receiver.
Specifically, the proposed G-ECCT resolves the mismatch between the inner and outer receiver by achieving symbol-level decoding.
Besides, the three-stage configuration-adaptive pre-training strategy enables joint optimization and generalization to diverse system configurations.
Extensive simulations demonstrate that the proposed receiver achieves superior performance across different system configurations.
It also exhibits strong zero-shot generalization capability to unseen frequency bands and channel models, while reducing inference latency compared with traditional iterative decoding methods.
These results indicate that the proposed unified architecture represents a step toward realizing AI-native receivers.
It provides a promising direction for building more efficient and intelligent physical-layer processing in future wireless systems.
Future research may explore more computationally efficient Transformer architectures~\cite{axis}, as well as model compression and quantization techniques~\cite{QAT} to further reduce the computational overhead and facilitate real-time deployment on computational resource-constrained hardware.

\footnotesize

\bibliographystyle{IEEEtran}
\bibliography{neural_receiver, IEEEabrv}

\normalsize

\end{document}